# Tunneling as a Source for Quantum Chaos

Ofir Flom,[1,2] Asher Yahalom[2], Haggai Zilberberg[1], L. P. Horwitz[1,3,4], Jacob Levitan[1],

[1]Physics Department, Ariel University, Ariel, Israel

[2]Department of Electrical & Electronic Engineering, Ariel University, Ariel, Israel

[3]Department of Physics, Bar Ilan University, Ramat Gan, Israsl

[4]School of Physics, Tel Aviv University, Ramat Aviv, Israel

## Abstract

We use an one dimensional model of a square barrier embedded in an infinite potential well to demonstrate that tunneling leads to a complex behavior of the wave function and that the degree of complexity may be quantified by use of the spatial entropy function defined by $S = -\int |\Psi(x,t)|^2 \ln|\Psi(x,t)|^2\, dx$. There is no classical counterpart to tunneling, but a decrease in the tunneling in a short time interval may be interpreted as an approach of a quantum system to a classical system. We show that changing the square barrier by increasing the height/width do not only decrease the tunneling but also slows down the rapid rise of the entropy function, indicating that the entropy growth is an essentially quantum effect.

## 1. Introduction

In order to find a quantum mechanical counterpart to the celebrated definition of classical chaos as being the exponential divergence of neighboring trajectories in phase space one may recognize the fast spreading of the wave function as an indication for chaotic behavior. This rests on the fact that the wave function may be thought of as representing an ensemble of points in phase space and that the spread with time is a measure for the divergence of those points. The claim that the wave function can be thought as representing an ensemble of trajectories is partly justified by the fact that the path integral includes an integration over all possible paths leading from one space time point to another and that an outcome of classical or quantum mechanical type is due to constructive or destructive interference of the amplitudes and a result of very delicate cancellations.

Ehrenfest's Theorem, that the mean position of a quantum state will follow a quasi-classical trajectory breaks down sooner (this time interval is called the "Ehrenfest time" [1]) for a chaotic system than for a regular one. In two papers by respectively Zaslavsky [1] and Ashkenazy et al [2] it was argued that the series expansion developed for the approach to the classical limit of a quantum evolution for any given small h (Planck's constant) fails for times larger than the Ehrenfest time. If the corresponding classical system is chaotic, the Ehrenfest time is smaller than for the case of a corresponding classical integrable system. The rapid spread and loss of the "Ehrenfest structure" of the wave packet, which we aim to show takes place in the presence of a tunneling barrier, have often been based as a prediction of a high degree of complexity in the resulting phenomena and eventually also as quantum chaos.

It is generally assumed that classical mechanics emerges from quantum mechanics by letting Planck's constant approach zero, but this procedure entails mathematical difficulties. Discussions of the classical limit have often been based on Ehrenfest's Theorem. Ballentine et al [3] have argued that the classical limit of a quantum state is an ensemble of classical orbits, not a single classical orbit. Moreover they concluded that Ehrenfest's Theorem is neither necessary nor sufficient to define the classical regime. The failure of the mean position in the quantum state to follow a classical orbit may still admit a centroid that approximately follows a classical orbit, and hence that a quantum state may behave "classical" even when Ehrenfest's Theorem does not apply.

The definition of quantum chaos as a fast spreading of the wave function has the advantage that it does not depend on the well-known definition of quantum chaos as emerging in a quantum system where the classical counterpart is chaotic. Such a definition is of limited use because not all examples of quantum systems which do not have any classical counterpart can be included by the definition. Moreover, we intend to show that exactly those systems including systems with barrier penetration often exhibit a high degree of complex behavior, and therefore naturally should be characterized as chaotic.

It has been argued that quantum chaos cannot exist due to the fact that the Schrödinger equation is linear and that the solution $\psi$ should not exhibit complex behavior. For example, it has been argued that one can represent the wave function by a linear combination over a complete set, for which the coefficients, due to the action of the Schrödinger equations, take on a time dependence that can be represented by a classical-like Hamiltonian model [4] that appears to provide linear equation with smooth behavior. Since such a system has an infinite number of degrees of freedom, However, the behavior could be highly complex [4]. Moreover, it is the square amplitude $|\psi|^2$ of $\psi$, which satisfies nonlinear equations coupled to the phase and determines the observed evolution of the system [5].

There have been many attempts to associate chaos with a change in entropy; a basic obstacle is that the evolution of the density matrix of a system is unitary, and that the von Neumann definition $Tr\rho\log\rho$ is therefore invariant. This theorem, however, does not apply to an entropy $S = -\int|\Psi(x,t)|^2 \ln|\Psi(x,t)|^2\,dx$ defined in terms of the local distribution $|\Psi(x,t)|^2$, on a segment along the real line. We find that an entropy defined in this way displays a rapid increase in the neighborhood of a barrier. These examples are based on phenomena which are intrinsically quantum in nature, such as tunneling and the behavior of wave functions in the proximity of a barrier.

Pattanayak and Schieve [6] studied a one dimensional problem with a Duffing potential, without external perturbation. They assumed that the quantum states maintain the form of squeezed coherent states, and found that the variables $\xi = \langle x \rangle$, $\eta = \langle \dot{x} \rangle$ , $\rho = \langle (x - \xi)^2 \rangle$, and $\pi = \dot{\rho}$ have equations of motion that can be generated by a classical Hamiltonian, and that these variables evolve chaotically. The assumption that the squeezed coherent state structure is maintained is rather restrictive. Nevertheless, they obtained highly complex behavior of the quantum state although this model is not chaotic in the classical limit [7]. We have chosen to study another quantum system without classical limit, such as tunneling through a barrier, in order to test if we obtain chaotic/complex behavior in systems without classical counterpart.

Tomsovic and Ullmo [8] used the Einstein-Brillouin-Keller method of quantization for a classical chaotic system and showed that the chaotic behavior in the classical system is correlated to the rate of tunneling in the quantum system. We shall take another route and show that even for classical systems without chaotic behavior the quantum mechanical analogue to the classical system may reveal chaotic-like behavior caused by tunneling and we shall investigate this mechanism in the model of a square barrier embedded in an infinite well.

The structure of this paper is as follows: first we describe the model including the parameters of the potential and describe the eigenfunctions and eigenvalues of the Hamiltonian. Then we study the tunneling probability and show how it is affected by the parameters of the potential. Following this we define the entropy of the wave function and study again the effect of the potential parameters. Finally we discuss the results and their implications for quantum chaos.

## 2. The Model

We use a one dimensional model of a square barrier embedded in an infinite potential well (see figure 1).

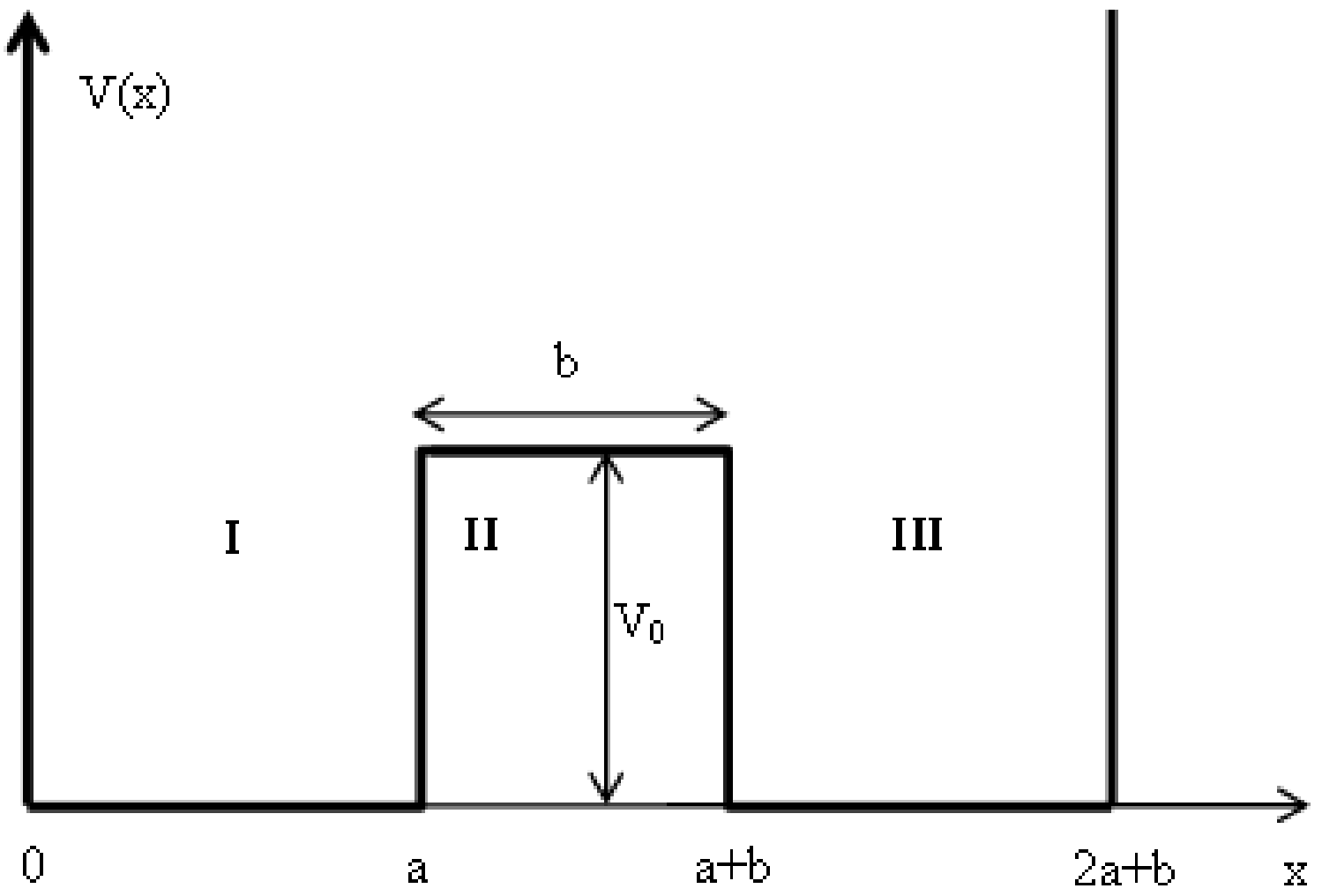

Fig 1 Symmetric double well

The wave packet is initially located on the LHS of the barrier. The half-width of the well is taken as a=35, the width of the Gaussian wave packet is σ=3, the barrier width is b and the barrier height is $V_0$. In all cases we use units such that $\hbar = 6.5821220 \cdot 10^{-16} eV \cdot sec$ and m=0.5109906MeV/$c^2$ which is the rest mass of an electron. The Gaussian wave packet is constructed from the first 30 energy levels. Those energy levels are obtained by matching the free wave functions in all the well parts such that the wave function and its derivatives are continuous. The eigenvalues for the Hamiltonian are given by the solutions of the following equation [9]:

$$\cosh(qb)\sin(2ka) + \frac{k^2 - q^2}{2qk}\sinh(qb)\cos(2ka) = -\frac{k^2 + q^2}{2qk}\sinh(qb) \qquad (1)$$

in which: $k = \sqrt{\frac{2m}{\hbar^2}E}$ and $q = \sqrt{\frac{2m}{\hbar^2}(V_0 - E)}$. The eigenfunction of the Hamiltonian are

$$\psi_n(x) = C_n \begin{cases} \sin(k_n x), 0 < x \le a \\ \sin(k_n a)\cosh[q_n(x-a)] + \frac{k_n}{q_n}\cos(k_n a)\sinh[q_n(x-a)], a < x \le a+b \\ \cos[k_n(x-a-b)]\left(\cosh(q_n b)\sin(k_n a) + \frac{k_n}{q_n}\sinh(q_n b)\cos(k_n a)\right) \\ + \sin[k_n(x-a-b)]\left[\cosh(q_n b)\cos(k_n a) + \frac{q_n}{k_n}\sinh(q_n b)\sin(k_n a)\right], a+b < x \le 2a+b \end{cases} \qquad (2)$$

where $C_n$ can be found by normalization $\int \left|\psi_n(x)\right|^2 dx = 1$.

We take the initial wave function to be:

$$\Psi(x,0) = \frac{1}{(2\pi\sigma^2)^{\frac{1}{4}}} e^{-\frac{(x-x_0)^2}{4\sigma^2}} e^{ik_0(x-x_0)} \qquad (3)$$

where we shall assume from now on $x_0$=11. If σ is small and $x_0$ inside the well, we need not take into account the infinite walls in specifying the initial wave function.

## 3. Probability for Tunneling

The wave function at all time is $\Psi(x,t)=\sum_{n=0}^{\infty}A_n\psi_n(x)e^{-i\frac{E_n}{\hbar}t}$, where $A_n=\langle\psi_n(x)|\Psi(x,0)\rangle$. The probability for tunneling is obtained from the expression [10]:

$$\rho_{RHS}(t)=\int_{a+b}^{2a+b}\Psi^*(x,t)\Psi(x,t)dx \qquad (4)$$

since

$$\Psi^*(x,t)\Psi(x,t)=\sum_n\sum_m A_m^*A_n\psi_m^*(x)\psi_n(x)e^{i\frac{E_m-E_n}{\hbar}t} \qquad (5)$$

If we use Eq. (5) in Eq. (4) we obtain

$$\rho_{RHS}(t)=\int_{a+b}^{2a+b}\sum_n\sum_m A_m^*A_n\psi_m^*(x)\psi_n(x)e^{i\frac{E_m-E_n}{\hbar}t}dx$$

$$=\sum_n\sum_m A_m^*A_ne^{i\frac{E_m-E_n}{\hbar}t}\int_{a+b}^{2a+b}\psi_m^*(x)\psi_n(x)dx \qquad (6)$$

Figure 2 describes the $\rho_{RHS}(t)$ as function of time for various heights of the barrier. Indeed one can see that the lower the barrier the higher the tunneling probability.

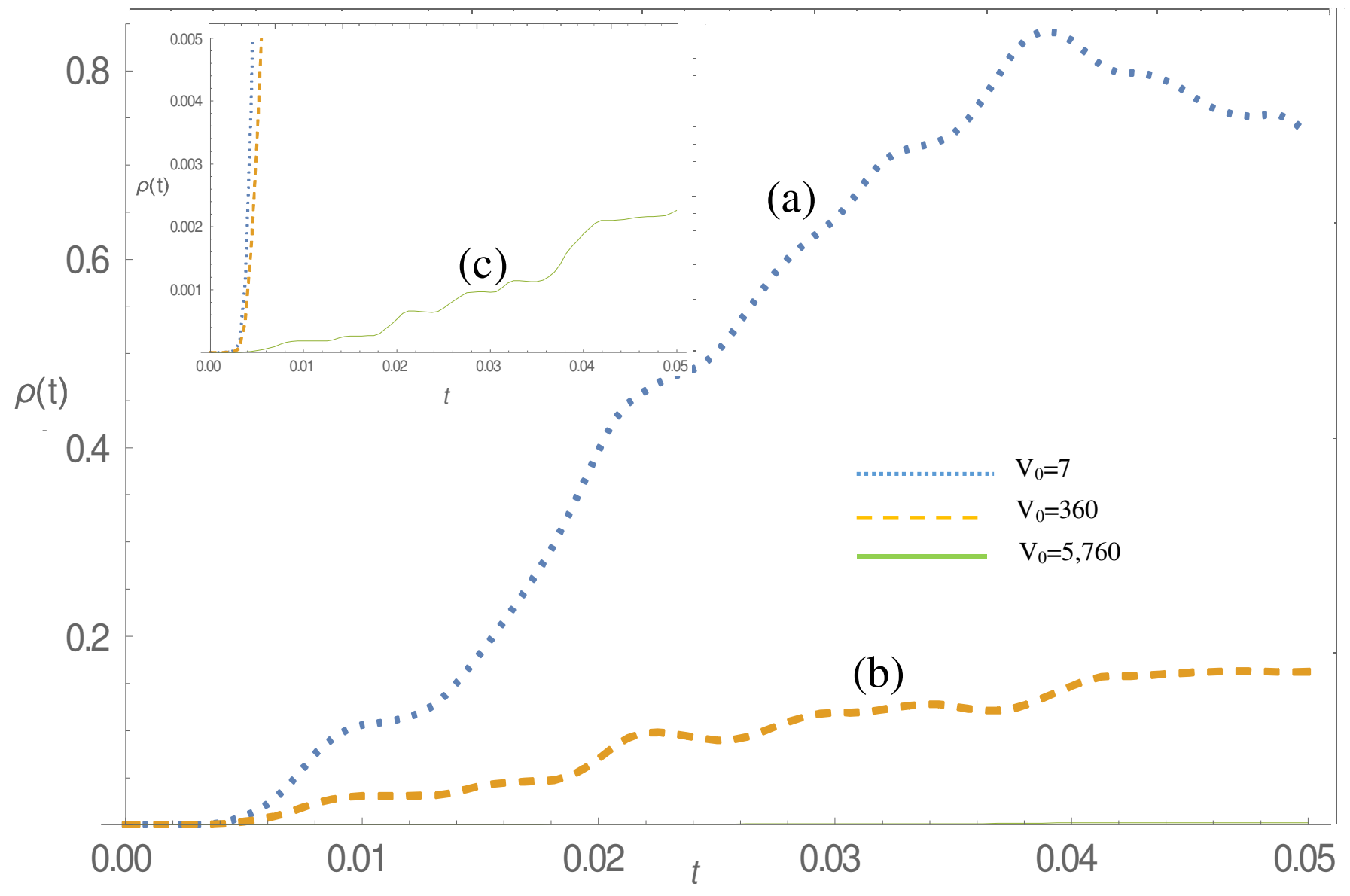

Fig 2. The tunneling probability as a function of time in the RHS of the well for three different heights of the barrier a) $V_0$=7, (b) $V_0$=360 and (c) $V_0$=5,760. The inset shows a change of scale of the tunneling probability for $V_0$=5,760.

## 4. Entropy

To demonstrate that tunneling leads to a complex behavior of the wave function and that the degree of complexity may be quantified by use of the spatial entropy function defined by

$$S(t) = -\int_{0}^{2a+b} |\Psi(x,t)|^2 \ln|\Psi(x,t)|^2 \, dx \qquad (7)$$

We study the way S changes as function of the parameters of the potential.

In fig 3 we see that changing the square barrier by increasing the height not only decreases the tunneling, but also slows down the rise of the entropy function. This demonstrates that increase of the barrier height reduces the complexity of the system, in accordance with the transition to a system which approaches the classical limit with no tunneling.

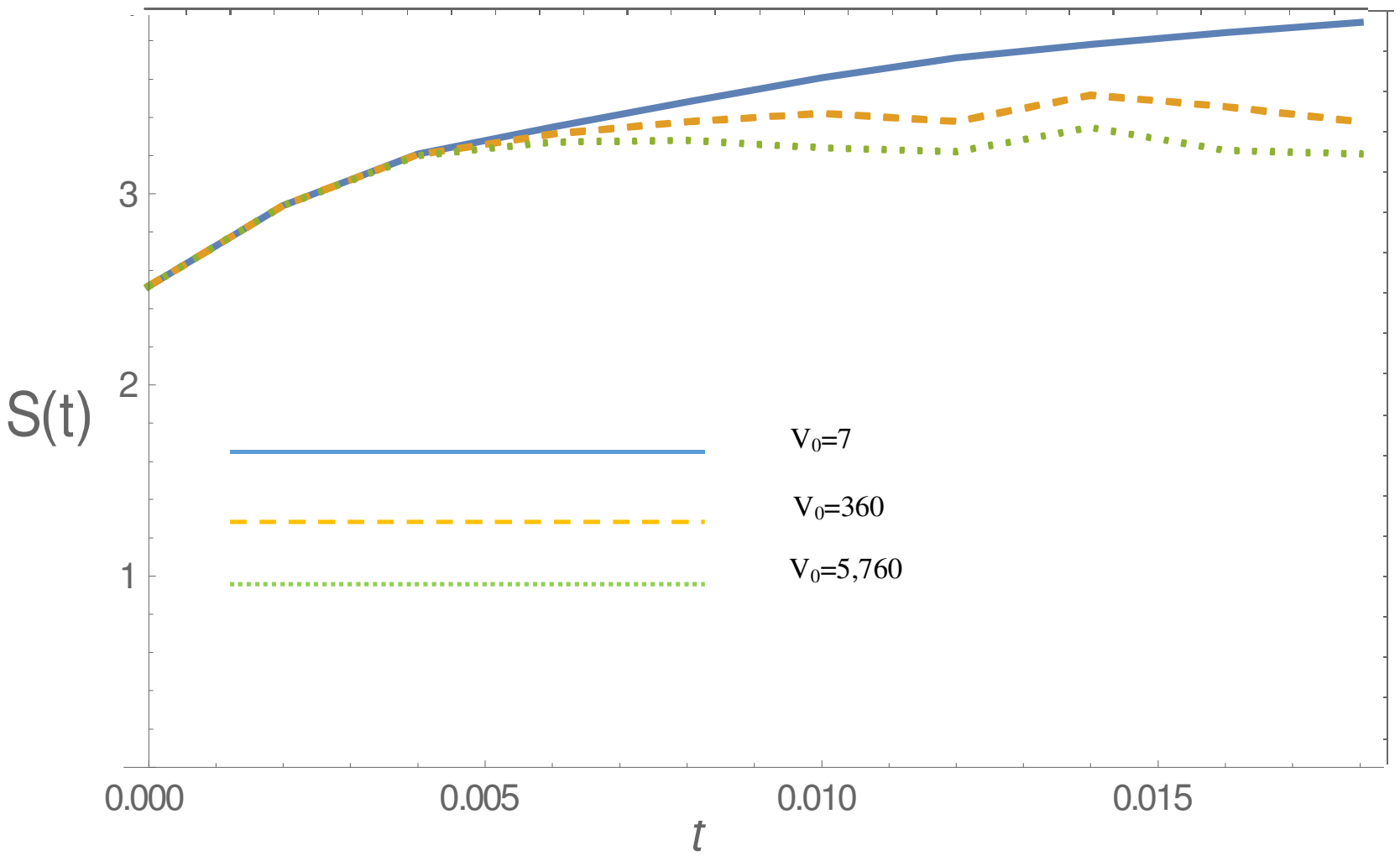

Figure 3 The spatial Entropy function for three different heights of the barrier.

In fig. 4 we see another example of reduced complexity/chaotic behavior in a transition to a system approaching the classical limit, but in this example the reduced tunneling which also causes reduced chaotic like behavior is obtained by increasing the width of the barrier.

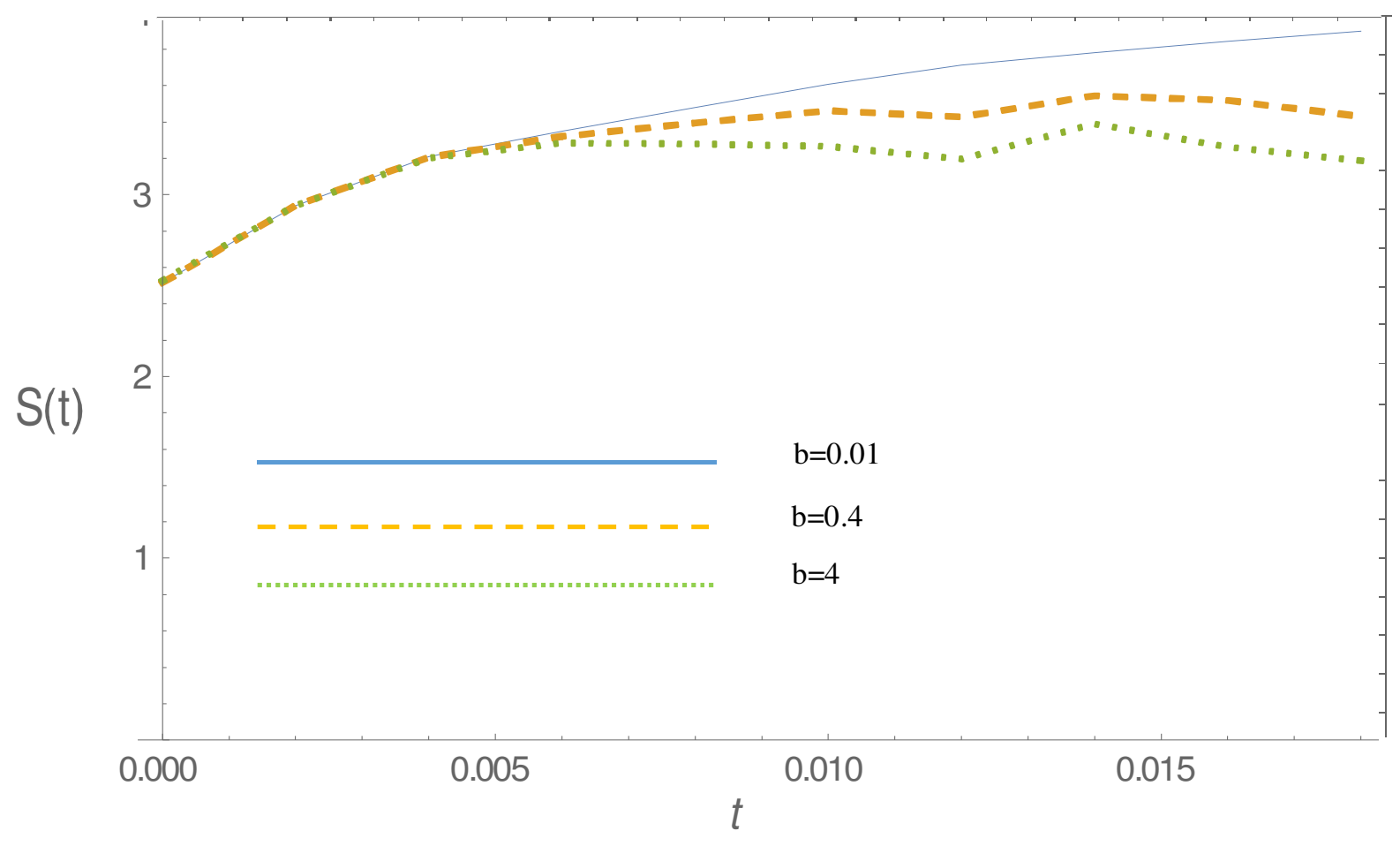

Figure 4 The spatial Entropy function for three different widths of the barrier.

In figure 5 and 6 we replace the spatial entropy function with the variance

$$\sigma^2(t) = \langle (x - \langle x \rangle)^2 \rangle \qquad (8)$$

which also "measures" the spread of the wave function in time. Figure 5 shows, similarly to figure 3 that increase/decrease in the barrier height reduces/enhances the tunneling which means reduces/enhances the chaotic like behavior.

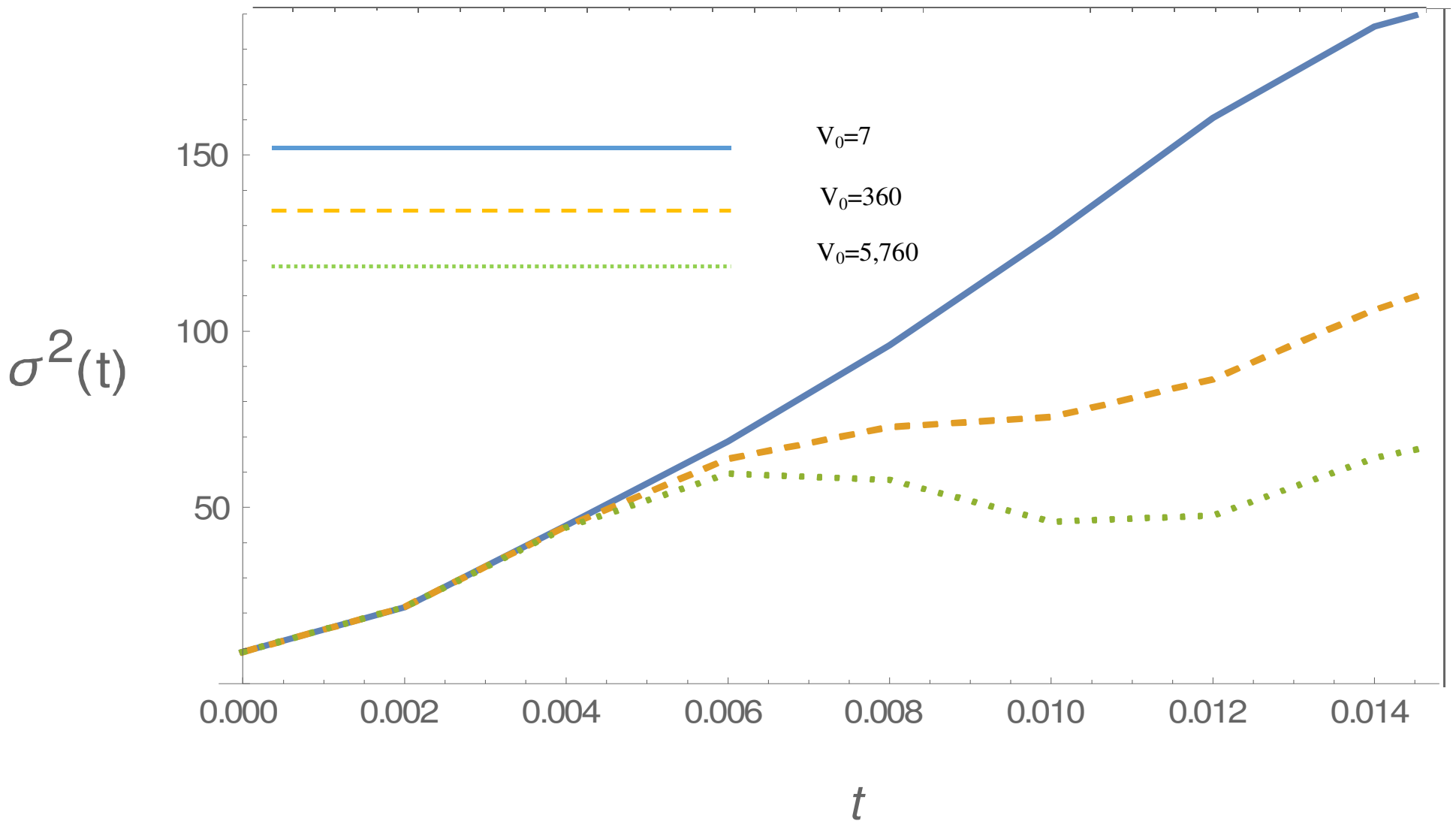

Figure 5 The Variances for three different heights of the barrier.

Fig 6 shows, similarly to fig 4, that increase/decrease in the width of the barrier reduces/enhances the tunneling and the chaotic like behavior. We conclude that the model in addition to not having any classical counterpart reveals increased chaotic behavior as the quantum system is removed further away from the classical limit.

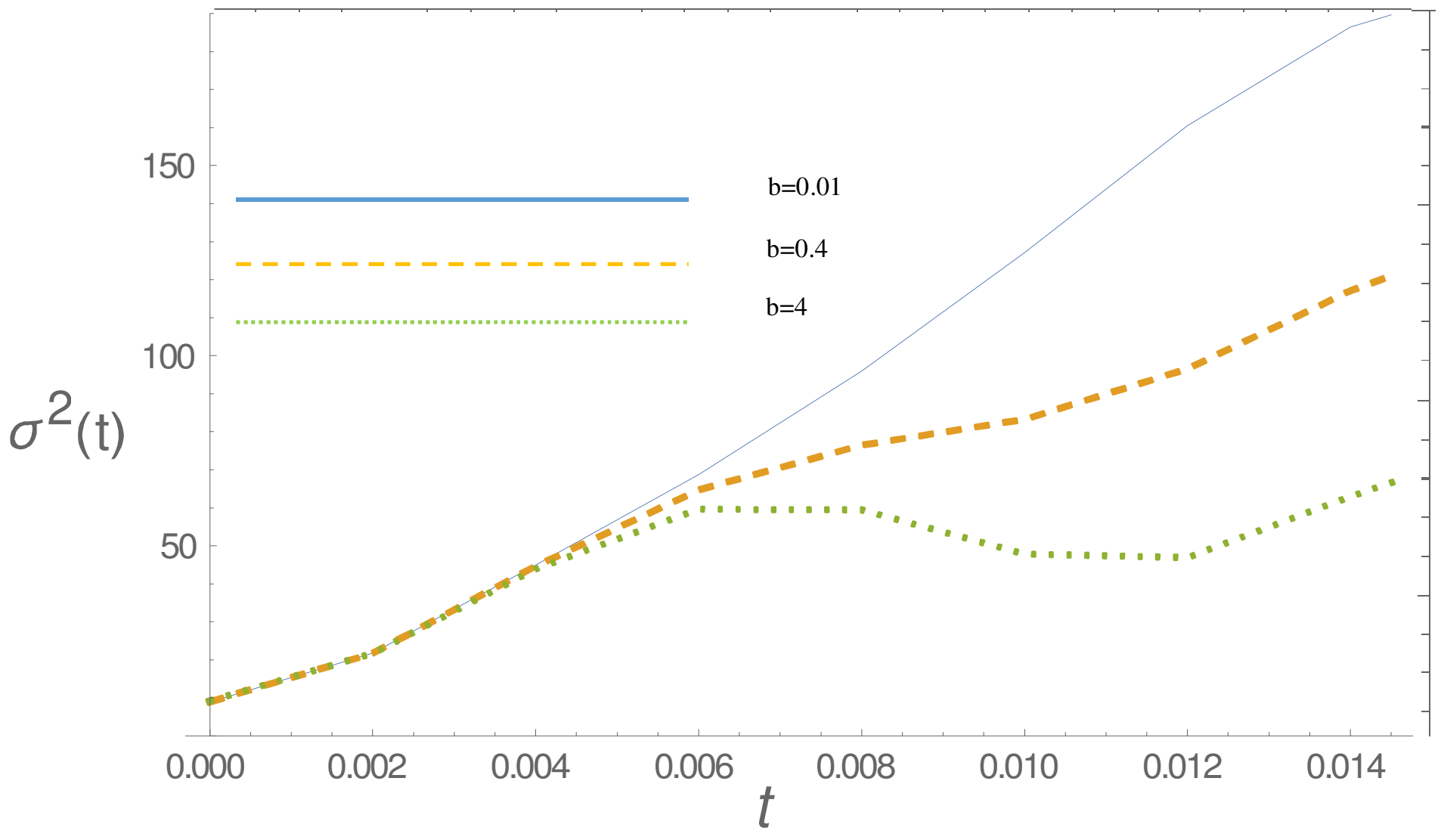

Figure 6 The Variances for three different widths of the barrier.

These results emphasize the relation between wave packet shape and the measure of complexity provided by the entropy.

## 5. Wave Function Evolution

In fig 7 we follow a Gaussian shaped classical probability density

$$\mathrm{P}_{cl}(x,t) = \frac{1}{\sqrt{2\pi\sigma^2}}\left(e^{\frac{-(x-x_0-vt)^2}{2\sigma^2}} + e^{\frac{-(x-2a+x_0+vt)^2}{2\sigma^2}}\right), \qquad x < a$$

and compare its time evolution with that of a quantum wave of identical shape. We compare the time evolution of the two waves for two different heights of the barrier: 360 and 5760. From figure 6 it appears that it takes 0.02 time units before the difference of the mean root square has grown from 0 to 125.58 in the case where the barrier height is 5760. For the barrier height 360 it takes only approximately the half of this time (0.0108) before the same difference of the mean root square is obtained.

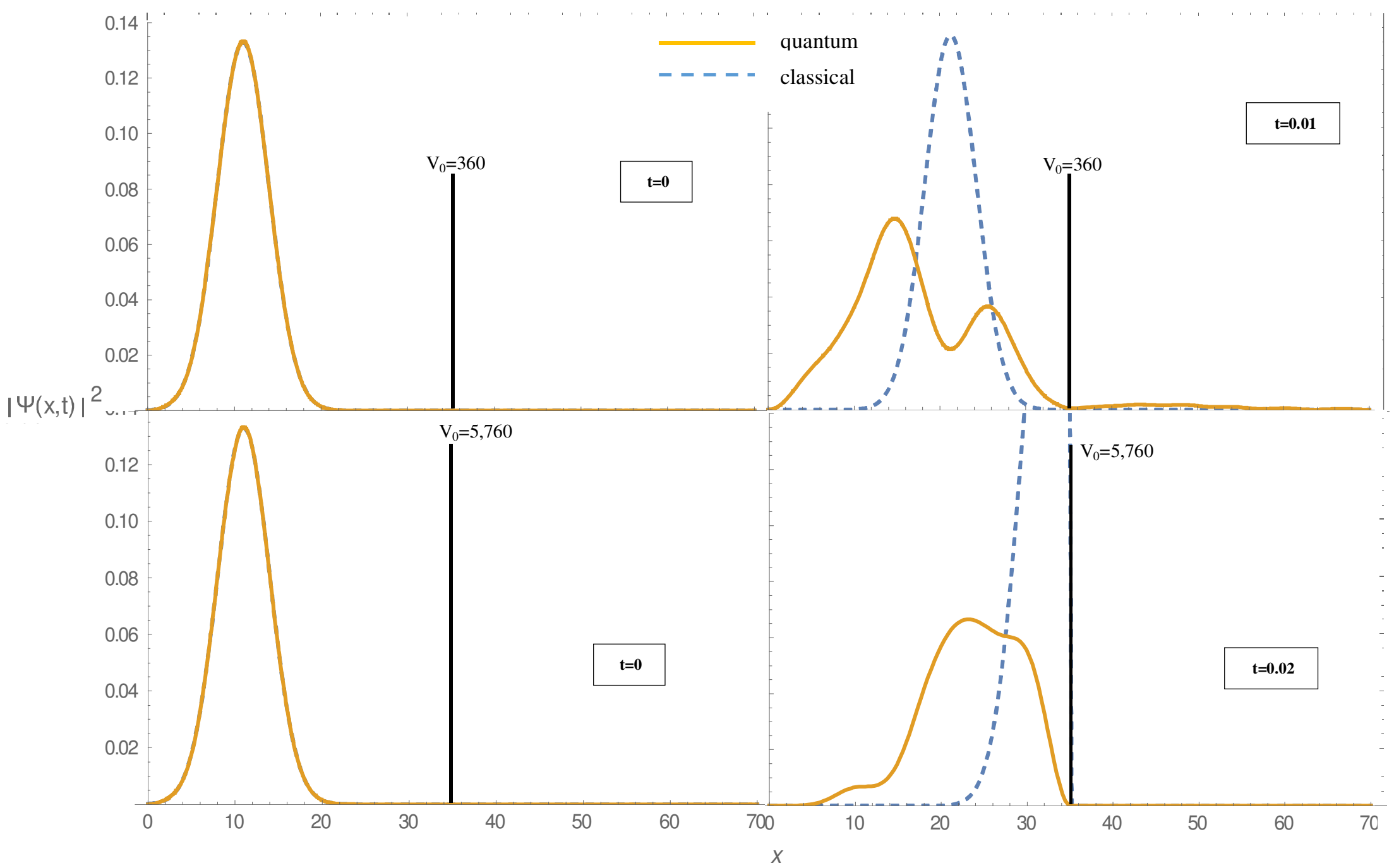

Fig 7. Comparison of two initially identical wave packets, one quantum mechanical and the other a classical probability density for two different barrier heights. for the barrier height $V_0$=5760 it takes approximately double as long time (0.0200) to reach a difference in the variance between the two wave packets equal to 125.58 as for the case with barrier height $V_0$ = 360 ( 0.0108)

We conclude that increase of the tunneling decreases the time it takes to obtain a certain difference in the position between the centroid of a classical probability density and that of a quantum wave which initially had the same position.

## 6. Instantons and tunneling

It is generally expected that asymmetric double well potentials allow for tunneling of a wave function initially placed in the "upper" well unto the "lower" well. However, Nieto et al. [10] showed that the tunneling from the false vacuum to the true vacuum is a very sensitive function of the shape of the potential.

The quantum mechanical treatment of the square barrier system lifts the unperturbed degeneracy and causes tunneling. It is well known that symmetric double-well potentials have ground state and first excited state almost degenerate [11]. Nieto et al. [10] maintain that the time it takes for a wave packet originally located on the one side to get to the other side is given by $t = \frac{\pi\hbar}{E_1 - E_0}$ and that the wave packet tunnels back and forth in multiples of this oscillation time.

In fact there is no consensus about what the tunneling time is but it is interesting to note that according to a rather general model proposed by Chiu, Sudarshan and Misra [12] the short time below which the survival probability decreases at most quadratically is given by

$$t = \frac{25\hbar}{E_{res} - E_{th}} \qquad (9)$$

where $E_{res}$ is the energy of the unperturbed state and $E_{th}$ the threshold energy, and Peres [13] has estimated the time for the onset of the exponential mode to be approximately

$$t = \frac{2\pi}{E_{res} - E_{th}} \qquad (10)$$

One may consider this estimate an indication of tunneling time, as in the alpha decay of a nucleus [14] not far from the estimates discussed by Landauer and Martin [15] and Davies [16]. The energy level splitting is

$$\mathrm{E}_{even} - \mathrm{E}_{odd} \approx \frac{4E_0\hbar}{b\sqrt{2mV_0}} e^{-\frac{b\sqrt{2mV_0}}{\hbar}} \qquad (11)$$

The action for an instanton is $S_0 = \int_a^{a+b} \sqrt{2mV(\phi)}d\phi = b\sqrt{2mV_0}$ which implies that the energy split for the ground state is proportional to $e^{-\frac{S_0}{\hbar}}$ and we conjecture that the energy split is related to a Euclidean path integral (see [17] for a more detailed discussion).
This is in accordance with our perception of the wave function as a collection of trajectories. The very fact that the path integral includes all possible paths from the initial to the final point, suggests that the wave function may be essentially associated with a set of potential trajectories.

The "birth" of the instanton is associated with the tunneling which can cause complexity in the evolution of the wave function signaling chaos characterized by an increase of the spatial entropy function (fig. 3 and 4). Delicate cancelations in the sum of the phases result in a relatively smooth form in the evolution of the wave function with highest degree of smoothness in the classical evolution and lesser in the quantum evolution where the tunneling causes the onset of decoherence together with a sharp increase of the entropy [18].

## 7. Barrier Displacement

It was shown in ref. [19] that by displacing the barrier in the double well, to the right or to the left, certain positions are passed where the system becomes somewhat almost degenerate.

These positions are distributed in almost commensurate intervals, and it is exactly for those positions that one may find significant tunneling accompanied by chaotic like behavior. When the width of the left well is twice the width of the right well or one obtains degeneracy for every second energy level. If the width of the left well is one third of that of the right well, every third level is degenerate. For the symmetric case (the center of the barrier placed in the center of the well) all levels are degenerate (a finite barrier splits the degeneracy [20]).

It was also shown, that in case of high-degeneracy, the tunneling has exponential decay for times not too small and not to large, while at other positions where almost degeneracy is somewhat weaker, the transition curve is non-exponential and develops oscillations. The rapid onset of exponential decay [12] may be associated with chaos in the sense that the spectral weight of the decay system should approach a Markovian distribution [21] and lead to irreversibility. Willox et al [22] showed that for a two-dimensional (classical) Sinai billiard a specific correlation function displays an initial non-exponential decay, and that the onset of the exponential decay corresponds to the onset of chaos in the system. The initial non-exponential era may be understood as the preparation time for the manifestation of chaos. Almost degeneracy enhances tunneling which facilitates chaos and vice versa. We therefore conjecture that systems with a high degree of almost degeneracy must have a sharp increase of the entropy.

Fig 8 shows the spatial entropy function for the barrier displaced to the right such that the left well is twice the width of the right well etc. Horwitz et al [23] showed that in cases of high-degeneracy the tunneling has exponential decay for times not too small and not too large, while at other positions, where almost degenerate conditions are weaker, the transition curve is non-exponential and develops strong oscillations before entering the exponential mode. For very large times the decay law again exhibits deviation from the exponential. Fig 8 reveals in accordance with our conjecture, that for locations with considerable almost degeneracy the entropy function has a sharp increase, while for locations where almost degenerate conditions are weaker, the entropy function has a slower increase.

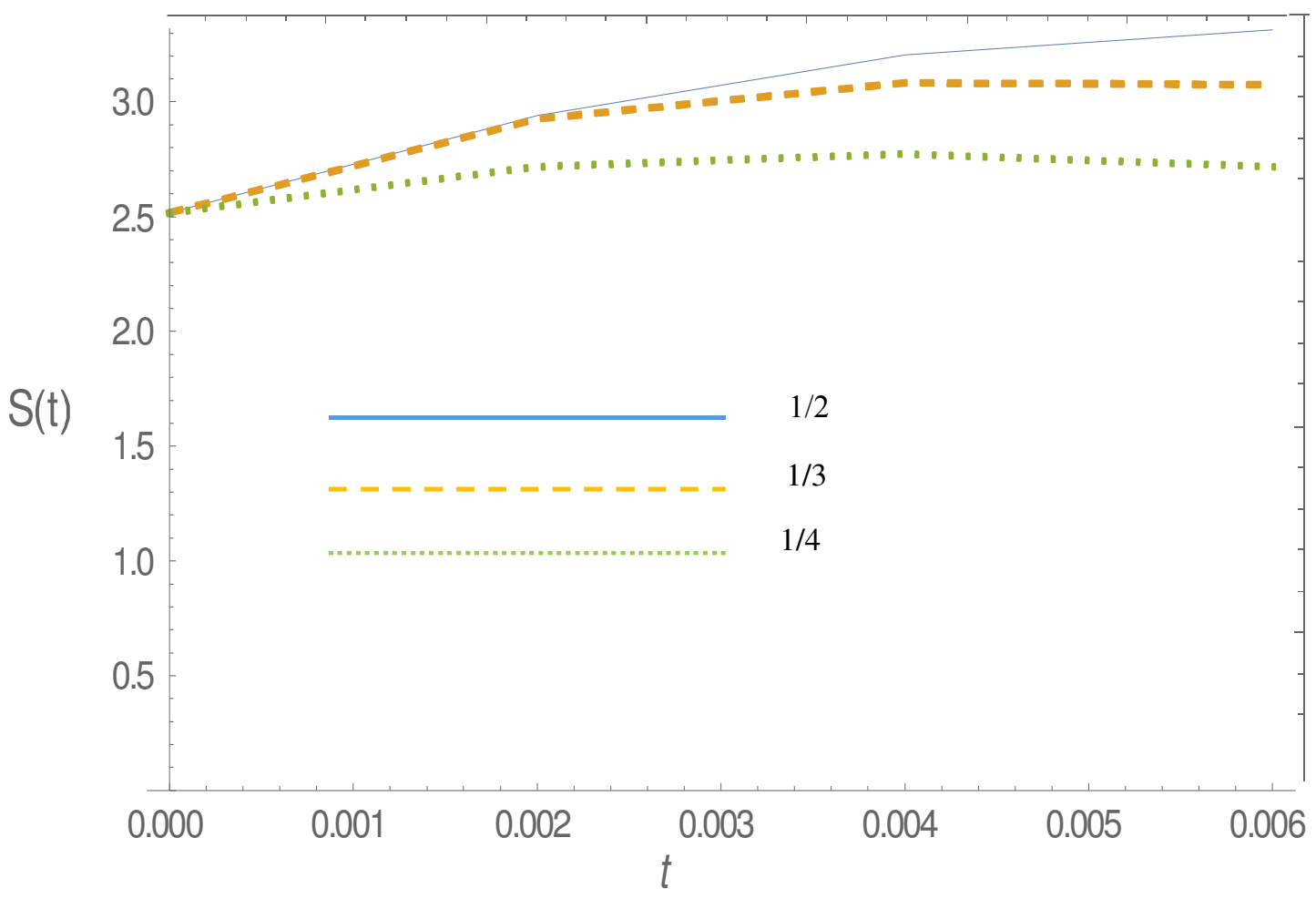

Figure 8 Graphs of Entropies for different positions of the barrier with height 360.

## 8. Ehrenfest time for chaotic systems

Ballentine et al. [3] concluded, that Ehrenfest's theorem breaks down much sooner for classical systems with chaotic motions than for classical systems with regular motions i.e. the Ehrenfest time is smaller for chaotic systems than for regular ones. We find an analogous result for quantum systems which have no classical counterpart. It can be stated that the Ehrenfest time is shorter for systems where the quantum system has a higher degree of chaotic behavior than for systems with a lesser degree of chaotic behavior, in which chaotic behavior is defined in terms of the entropy function.

In addition to the fact that chaotic behavior accompanies tunneling, and taking into account that tunneling was found to be enhanced by a high degree of almost degeneracy we conclude, that quantum systems with a higher degree of complex, chaotic-like behavior, have a sharper increase of entropy, than systems with lower degeneracy. It was furthermore shown [23] that the time scale on which decoherence takes place depends on the degree of complexity of the underlying quantum mechanical system, i.e., more complex systems decohere faster than less complex systems [24].

In this work, we have seen that a system in a stable, ordered state may evolve, in the presence of tunneling to a chaotic state associated with an increase in entropy.